\documentclass[preprint,amsmath,amssymb,longbibliography,showkeys]{revtex4}

\usepackage{graphicx}
\usepackage{color}
\usepackage{dcolumn}
\usepackage{bm}
\usepackage{epstopdf}
\usepackage{hyperref}

	\usepackage{fancyheadings}
\renewcommand{\thispagestyle}[1]{} 

\begin{document}

\title{Thermodynamic properties of the one-dimensional Ising model\\
 with magnetoelastic interaction}

\author{T. Balcerzak}
\email{tadeusz.balcerzak@uni.lodz.pl}
\homepage[]{https://orcid.org/0000-0001-7267-992X}
\author{K. Sza{\l}owski}
\email{karol.szalowski@uni.lodz.pl}
\homepage[]{https://orcid.org/0000-0002-3204-1849}
\affiliation{%
University of \L\'{o}d\'{z}, Faculty of Physics and Applied Informatics,\\
Department of Solid State Physics, ulica Pomorska 149/153, PL90-236 \L\'{o}d\'{z}, Poland
}%

\author{M. Ja\v{s}\v{c}ur}
\email{michal.jascur@upjs.sk}
\homepage[]{https://orcid.org/0000-0003-0826-1961}
\affiliation{%
Department of Theoretical Physics and Astrophysics, Faculty of Science,\\
P. J. \v{S}af\'arik University, Park Angelinum 9, 041 54 Ko\v{s}ice, Slovak Republic
}%


\begin{abstract}
The Ising one-dimensional (1D) chain with spin $S=1/2$ and magnetoelastic interactions is studied with the lattice contribution included in the form of elastic interaction and thermal vibrations simultaneously taken into account. The magnetic energy term and the elastic (static) energy term based on the Morse potential are calculated exactly. The vibrational energy is calculated in the Debye approximation, in which the anharmonicity is introduced by the Gr{\"u}neisen parameter. The total Gibbs potential, including both the magnetic field, as well as the external force term, is constructed and from its minimum the equation of state is derived.

From the Gibbs energy all the thermodynamic properties are calculated in a self-consistent manner. The comprehensive numerical calculations are performed in a full temperature range, i.e., from zero temperature up to the vicinity of melting. In particular, a role of magneto-elastic coupling is emphasized and examined. The numerical results are illustrated in figures and discussed. 

\end{abstract}

\keywords{Ising model; Magnetoelastic coupling; Thermodynamics of magnets; Thermodynamic response functions; Magnetostriction}
\maketitle

\newpage

\section{Introduction}

The one-dimensional (1D) Ising model \cite{Ising1925} plays an important role in the theory of magnetism, being one of the models which have been solved exactly \cite{Baxter1982, Salinas2001, Strecka2015b}. The generalized versions of this model have been applied to higher dimensions, different lattices or modified magnetic interactions. As far as one dimension is concerned, the model has been extended to the general spin value $S>1/2$ \cite{Obakata1968, Jascur1992}, magnetic long-range \cite{Dobson1969, Nagle1970, Mejdani1994, Juhasz2014} and multi-spin interactions \cite{Fan2011, Turban2016}. The various quantum generalizations of 1D model have also been studied \cite{Crisanti1994, Ovchinnikov2003, Damski2015}.

Apart from the interest in the Ising model due to its significance in statistical physics, some of its importance is connected with its application for the description of quasi-1D magnetic systems \cite{Wolf}. In such context, systems such as CoCl$_2 \cdot$ 2NC$_5$H$_5$, CoCl$_2 \cdot$ 2H$_2$O, (CH$_3$)$_3$NHCoCl$_3 \cdot$ 2H$_2$O \cite{Jongh1974a, Steiner1976}, CoCl$_2 \cdot$ 2D$_2$O \cite{Larsen2017}, or BaCo$_2$V$_2$O$_8$ \cite{Faure2018a} can be mentioned. In another class of quasi-1D materials, i.e., in spin-crossover systems like Fe-based chain compounds \cite{Linares1999}, or  copper-based chain polymer heterospin complexes \cite{Morozov2010}, as a result of the deformation of the spin-changing molecules, the elastic long-range couplings have been taken into account in addition to the magnetic interaction \cite{Boukheddaden2000, Boukheddaden2007}. An application to 1D ferroelectric chains, like Ca$_3$CoMnO$_6$ \cite{Yao2008, Yao2009, Yun-Jun2009, Qi2011}, is also known, with the extension to studies of the magnetocaloric effect \cite{Qi2016}.
It is worth mentioning that the model of Ising chain has also been applied to the statistical genetics \cite{Majewski2001, Colliva2015}, including DNA high-force stretching \cite{Storm2003} and to the chiral homopolymers description \cite{Chernodub2011}.

It has been known that by including elastic interactions the thermodynamic properties of the 1D Ising model can be markedly influenced \cite{Enting1973, Salinas1974, Figueiredo1978}. Such extension gained considerable attention and has been carried out both from the point of view of purely model research \cite{Mijatovic1977, Djordjevic1978, Mijatovic1980, Knezevic1980, Axel1981, Timm2006, Barsan2010, Lemos2019}, and the implementation for particular experimental systems as well \cite{Boukheddaden2007, Yun-Jun2009, Morozov2010}. In most of these papers the atoms forming 1D chain are treated as coupled harmonic oscillators \cite{Salinas1974, Axel1981, Yun-Jun2009}. Rarely, the oscillators with quartic anharmonicities have been considered \cite{Barsan2010}. The magneto-elastic couplings are taken into account via simplified, linear dependency of the exchange integral on the interatomic distance \cite{Salinas1974, Figueiredo1978, Mijatovic1980, Morozov2010}. According to our knowledge, in one case \cite{Yun-Jun2009} the exchange integral has been assumed in the form of Lennard-Jones potential. However, in spite of such intensive studies, there is still lack of complete thermodynamic theory which would be able to describe 1D system in a self-consistent way, simultaneously taking into account the magnetic, elastic (static) and vibrational (thermal) properties.

In our proposition of the thermodynamic description we start from the Gibbs energy construction, which consists of the magnetic (Ising) part, the elastic (static) energy, the vibrational (thermal) energy and the external force term. Such a method has already been implemented for the description of the bulk models with magneto-elastic interactions \cite{Balcerzak2010a, Balcerzak2017b, Balcerzak2018c}. Recently, the method has also been adopted for the Hubbard pair-cluster with elastic inter-atomic potential in the external fields \cite{Balcerzak2019}. 

In the case of 1D Ising chain the magnetic energy can be calculated exactly in the presence of the external magnetic field. The elastic (static) energy is taken in a form resulting from the Morse potential, which is anharmonic and in the case of linear chain can also be calculated exactly. The coupling between the magnetic and elastic terms is introduced by the nonlinear, power-law dependence of the exchange integral vs. inter-atomic distance.
On the other hand, the vibrational (thermal) energy is approximated by the extended Debye model, in which the anharmonicity of the Morse potential is taken into account via the Gr{\"u}neisen parameter. This parameter has been known exactly and it assures the consistency of the description of static and vibrational energies, both based on the Morse potential.
The total Gibbs potential, including also the external force term, is then minimized with respect to the inter-atomic distance deformation, which leads to the equation of state (EOS) and describes thermodynamic equilibrium.

The paper is organized as follows: In the next section the theoretical method is presented in detail. In the subsequent section the numerical results are illustrated in figures and their discussion is given. The last section includes summary and final conclusions.\\

\section{\label{sec:theory}Theoretical model}

The present section contains a detailed description of the Ising model extended by taking into consideration the elastic and vibrational terms. The equation of state is derived for the model in question and the fundamental thermodynamic quantities are calculated. 

\subsection{Magnetic energy}

Magnetic subsystem of 1D chain is described by the Ising Hamiltonian:
\begin{equation}
\mathcal{H}_{\rm I} = -J\sum_{\left< i,j \right>}^{N}S_i S_j -h\sum_{i}^{N}S_i,
\label{eq1}
\end{equation}
where the $z$-component of the spin on $i$-site takes the values $S_i = \pm 1/2$. $J \ge 0$ is the ferromagnetic exchange integral, limited to nearest neighbours (NN), and $h$ stands for the external magnetic field. It is known that such a model has been solved exactly, for instance, by the transfer matrix method and the magnetic Gibbs energy per spin has been found in the form of \cite{Salinas2001}:
\begin{equation}
\frac{G_{\rm I}}{N}=-\frac{1}{4}J-k_{\rm B}T \ln \left[\cosh \left(\frac{\beta h}{2}\right)
+\sqrt{\cosh^2 \left(\frac{\beta h}{2}\right) +e^{-\beta J} -1}\;\right],
\label{eq2}
\end{equation}
where $\beta = 1/k_{\rm B}T$. Now, in order to take into account the magneto-elastic effects, we introduce the power-law dependence of the exchange integral vs. NN distance $d$, namely:
\begin{equation}
J=J_0\left(\frac{d}{d_0}\right)^{-n}=J_0\left(1+\varepsilon\right)^{-n}.
\label{eq3}
\end{equation}
It should be mentioned that the power-law dependence for the exchange integral on the interatomic distance has been confirmed by many experimental studies showing good fit with the experimental data 
\cite{Rogers_1972, Denissen_1986, Szuszkiewicz_2006}.
In Eq.(\ref{eq3}) we used the relation:
\begin{equation}
d=d_0\left(1+\varepsilon\right),
\label{eq4}
\end{equation}
defining $d_0$ as the equilibrium distance between NN in the ground state, and $\varepsilon$ stands for the small relative change of the interatomic distance ($\varepsilon \ll 1$). We assume that $\varepsilon =0$ for $T=0$, when the magnetic field is absent ($h=0$) and no external force is applied ($f_s=0$). The exponent $n > 0$ in Eq.(\ref{eq3}) is a parameter which should assure the quick damping of $J$ vs. the distance, in agreement with the fact that only NN interactions are relevant, whereas the deformation $\varepsilon$ is small.

\subsection{Elastic crystalline energy}

The elastic (static) energy in 1D system can be conveniently found on the basis of the Morse pair potential \cite{Morse1929a}:
\begin{equation}
U\left(r_k\right)=D\left[1-e^{- \delta\left(r_k-r_0\right)/r_0}\right]^2.
\label{eq5}
\end{equation}
It is worth noticing that the Morse potential has been used to describe interatomic interactions in many crystalline 
metals \cite{Girifalco_1959, Lincoln_1967}.
In Eq.(\ref{eq5}) $D$ is the potential depth, $r_0$ corresponds to the equilibrium distance between two atoms forming an isolated pair when only the elastic energy is taken into account, and $\delta$ describes the potential width and its asymmetry. The inter-atomic distance $r_k$ between atoms being $k$-th neighbours can be expressed as:
\begin{equation}
r_k=kd=kd_0\left(1+\varepsilon\right), \;\;\;\;\;\; \left(k=1,2,\dots \right).
\label{eq6}
\end{equation}
The elastic energy per atom can be obtained by performing summation over all the distances (all the pairs):
\begin{equation}
\frac{U_{\varepsilon}}{N}=D\sum_{k=1}^{\infty}\left \{\left[1-e^{- \delta\left(k\frac{d_0}{r_0}\left(1+\varepsilon \right)-1\right)}\right]^2 - \left[1-e^{- \delta\left(k\frac{d_0}{r_0}-1\right)}\right]^2\right \}.
\label{eq7}
\end{equation}
In Eq.(\ref{eq7}) the elastic energy is normalized by the requirement that $U_{\varepsilon}=0$ for $\varepsilon=0$. i.e., it vanishes in a non-deformed state. One can notice that summation in Eq.(\ref{eq7}) can be performed exactly in 1D system using the formula for the sum of geometric series. Namely, by introducing the abbreviate notation:
\begin{equation}
E=e^{- \delta \frac{d_0}{r_0}\left(1+\varepsilon \right)}
\label{eq8}
\end{equation}
and
\begin{equation}
E_0=e^{- \delta\frac{d_0}{r_0}},
\label{eq9}
\end{equation}
we obtain the result in the form of:
\begin{equation}
\frac{U_{\varepsilon}}{N}=2De^{\delta}\left(\frac{E_0}{1-E_0}-\frac{E}{1-E}\right) + 
De^{2\delta}\left(\frac{E^2}{1-E^2}-\frac{E_0^2}{1-E_0^2}\right).
\label{eq10}
\end{equation}

\subsection{Vibrational energy}

The vibrational energy of the system in question is calculated within Debye approximation, in which the thermal excitations can propagate along the chain, whereas each atom is treated as three-dimensional oscillator. 
For such model, the Helmholtz free-energy  is calculated from a general formula \cite{Solyom2007}:
\begin{equation}
F_{\rm D}=3k_{\rm B}T \int_{0}^{\omega_{\rm D}} \ln \left[ 2\sinh \left( \frac{\hbar \omega}{2k_{\rm B}T }\right) \right] D\left(\omega \right) d\omega,
\label{eq10b}
\end{equation}
where $D\left(\omega \right)$ presents the density of states, and $\omega_{\rm D}$ is a Debye cut-off frequency.
We note that for $D\left(\omega \right)=N \delta \left(\omega - \omega_{\rm E}\right)$ the integral is trivial, and the formula (\ref{eq10b}) reduces to the free-energy of Einstein model with a single frequency $\omega_{\rm E}$. However, for 1D system in the Debye approximation the density of states takes a form $D\left(\omega \right)=N/ \omega_{\rm D}$, in agreement with Ref.~\cite{Solyom2007} (see eq. 12.2.22 for $L=1$ in Ref.~\cite{Solyom2007}). As a result, the Helmholtz free-energy per atom in the Debye approximation can be expressed as:
\begin{equation}
\frac{F_{\rm D}}{N}=\frac{3}{4}k_{\rm B}T_{\rm D}+3k_{\rm B}T \ln\left(1-e^{-x_{\rm D}}\right)-3k_{\rm B}T
\frac{1}{x_{\rm D}}\int_{0}^{x_{\rm D}} \frac{x}{e^x-1}dx,
\label{eq11}
\end{equation}
where $x_{\rm D}=T_{\rm D}/T$, and $T_{\rm D}$ is the Debye temperature 
defined by the relation $k_{\rm B}T_{\rm D}=\hbar \omega_{\rm D}$. Now, we take into account the fact that the Debye temperature depends on the system length $L$, thus on its relative deformation $\varepsilon$. Namely, according to the idea of Gr{\"u}neisen \cite{Grueneisen1912}, and following the approach presented in Ref. \cite{Balcerzak2010a} for the frequencies of anharmonic oscillators, we assume that $\omega_{\rm D} \propto 1/L^{\gamma}$, where $L=Nd=Nd_0\left(1+\varepsilon\right)$ is the length of the chain, and  $\gamma$ is the Gr{\"u}neisen parameter \cite{Grueneisen1912}. Thus, making use of the proportionality $\omega_{\rm D} \propto T_{\rm D}$ we can write:
\begin{equation}
T_{\rm D}=\frac{T_{\rm D}^0}{\left(1+\varepsilon\right)^{\gamma}},
\label{eq12}
\end{equation}
where 
$T_{\rm D}^0$ is the Debye temperature in the ground state (for $T=0$, $h=0$, and $f_s=0$).
On the basis of Eq.(\ref{eq12}) one can easily check that:
\begin{equation}
\gamma=-\frac{\left(1+\varepsilon \right)}{T_{\rm D}}\frac{\partial T_{\rm D}}{\partial \varepsilon}=-\frac{L}{T_{\rm D}}\frac{\partial T_{\rm D}}{\partial L},
\label{eq13}
\end{equation}
i.e., for 1D system the $\gamma$-parameter satisfies the equation involving the system length $L$, which is analogous to that postulated by Gr{\"u}neisen \cite{Grueneisen1912} for the system volume $V$.

The Debye integral in Eq.(\ref{eq11}) can be calculated exactly with the help of polylogarithmic functions \cite{Balcerzak2017b}. Namely,
\begin{equation}
\int_{0}^{x_{\rm D}} \frac{x}{e^x-1}dx = \frac{\pi^2}{6}+x_{\rm D}\ln\left(1-e^{-x_{\rm D}}\right)-
{\rm Li}_2\left(e^{-x_{\rm D}}\right),
\label{eq14}
\end{equation} 
where the second order polylogarithm, ${\rm Li}_2\left(e^{-x_{\rm D}}\right)$, is given by:
\begin{equation}
{\rm Li}_2\left(e^{-x_{\rm D}}\right)=\sum_{k=1}^{\infty}\frac{e^{-kx_{\rm D}}}{k^2}.
\label{eq15}
\end{equation}
In this way, the vibrational (thermal) energy (Eq. \ref{eq11}) of 1D chain can be presented in the following final form:
\begin{equation}
\frac{F_{\rm D}}{N}=3k_{\rm B}T_{\rm D}\left \{  \frac{1}{4}+\frac{1}{x_{\rm D}^2}\left[{\rm Li}_2\left(e^{-x_{\rm D}}\right)-\frac{\pi^2}{6} \right] \right \}.
\label{eq16}
\end{equation}

\subsection{Equation of state}

The total free energy for 1D chain is constructed as a sum of all the following contributions:
\begin{equation}
G=G_{\rm I}+U_{\varepsilon}+F_{\rm D}+f_sNd_0\left(1+ \varepsilon\right),
\label{eq17}
\end{equation}
where the magnetic, elastic and vibrational energies are given by the equations (\ref{eq2}),  (\ref{eq10}) and (\ref{eq16}), respectively. The last term in Eq. (\ref{eq17}) corresponds to the mechanical (enthalpic) part introduced by the external force $f_s$. We assume the convention in which $f_s>0$ corresponds to the compressive force, whereas $f_s<0$ stands for the stretching one. We note that the linear deformation $\varepsilon$ occurs in every term of right-hand-side of Eq. (\ref{eq17}), hence it can be treated as a variational parameter. In equilibrium, this parameter must minimize the total free energy at arbitrary temperature $T$, external magnetic field $h$ and force $f_s$. Thus, we demand that:
\begin{equation}
\frac{1}{N}\left(\frac{\partial G}{\partial \varepsilon}\right)_{T, h, f_s}=0.
\label{eq18}
\end{equation}
Eq. (\ref{eq18}) is equivalent to the balance of all forces in the system, namely we get from it:
\begin{equation}
f_{\rm I}+f_{\varepsilon}+f_{\rm D}=f_s,
\label{eq19}
\end{equation}
where $f_{\rm I}$, $f_{\varepsilon}$ and $f_{\rm D}$ are the magnetic (Ising), elastic (static) and thermal vibrational forces, respectively. These forces can be expressed in dimensionless units and are given by the formulas:
\begin{equation}
\frac{d_0f_{\rm I}}{J_0}=\frac{n}{\left(1+\varepsilon \right)^{n+1}}\left \{\frac{1}{2}
\frac{e^{-\beta J}}{\cosh \left(\frac{\beta h}{2}\right)
\sqrt{\cosh^2 \left(\frac{\beta h}{2}\right) +e^{-\beta J} -1}+\cosh^2 \left(\frac{\beta h}{2}\right) +e^{-\beta J} -1} -\frac{1}{4} \right \},
\label{eq20}
\end{equation}

\begin{equation}
\frac{d_0f_{\varepsilon}}{J_0}=-2\frac{D}{J_0}\frac{d_0}{r_0}\delta e^{\delta}\left[
\frac{E}{\left(1-E \right)^2} -e^{\delta}\frac{E^2}{\left(1-E^2 \right)^2} \right],
\label{eq21}
\end{equation}
and
\begin{eqnarray}
\frac{d_0f_{\rm D}}{J_0}&=&3\left(\frac{k_{\rm B}T_{\rm D}^0}{J_0}\right)\frac{\gamma}{\left(1+\varepsilon \right)^{\gamma+1}}\left \{\frac{1}{4}+\frac{1}{x_{\rm D}^2}\left[{\rm Li}_2\left(e^{-x_{\rm D}}\right)-\frac{\pi^2}{6} \right]\right \}\nonumber\\
&-&3{\left(\frac{k_{\rm B}T_{\rm D}^0}{J_0}\right)\!}^2\left(\frac{J_0}{k_{\rm B}T}\right)\frac{\gamma}{\left(1+\varepsilon \right)^{2\gamma+1}}\frac{1}{x_{\rm D}^2}
\left \{\frac{2}{x_{\rm D}}\left[{\rm Li}_2\left(e^{-x_{\rm D}}\right)-\frac{\pi^2}{6} \right] - \ln \left(1-e^{-x_{\rm D}} \right)\right \}.
\label{eq22}
\end{eqnarray}
In derivation of the last equation (Eq. \ref{eq22}) we made use of the identity:
\begin{equation}
\frac{\partial}{\partial \varepsilon}{\rm Li}_2\left(e^{-x_{\rm D}}\right)=
-\frac{\partial x_{\rm D}}{\partial \varepsilon}{\rm Li}_1\left(e^{-x_{\rm D}}\right)=
\frac{\partial x_{\rm D}}{\partial \varepsilon}\ln \left(1-e^{-x_{\rm D}} \right).
\label{eq23}
\end{equation}
Eq. (\ref{eq19}) presents the Equation of State (EOS) from which the linear deformation $\varepsilon$ can be found for given $T$, $h$ and $f_s$. However, prior to its usage, the constant parameter $d_0/r_0$, describing the non-deformed NN distance, should be determined. This parameter should be calculated from Eq. (\ref{eq19}), in which we put the conditions corresponding to the non-deformed ground state, i.e., $\varepsilon =0$, $T=0$, $h=0$ and $f_s=0$.

With the help of EOS the magnetization per spin, $m=\left<S_i\right>$, can be found from the formula:
\begin{equation}
m=-\frac{1}{N}\left(\frac{\partial G}{\partial h}\right)_{T,f_s},
\label{eq24}
\end{equation}
where $G$ is given by Eq. (\ref{eq17}). As we mentioned before, all terms in Eq. (\ref{eq17}) depend on $\varepsilon $, thus the derivative $\left(\partial \varepsilon/\partial h\right)_{T,f_s}$ must be taken into account in all these expressions. However, with the use of EOS (Eq. (\ref{eq19})) all contributions containing such derivatives will cancel, and the final result can be presented in the simple form:
\begin{equation}
m=\frac{1}{2}\frac{\sinh \left(\frac{\beta h}{2} \right)}{\sqrt{\sinh^2 \left(\frac{\beta h}{2} \right)
+e^{-\beta J}}},
\label{eq25}
\end{equation}
where $J(\varepsilon)$ is given by Eq. (\ref{eq3}) and $\varepsilon$ is obtained from EOS. It is worth noticing that for $\varepsilon =0$ we obtain $J=J_0$ and the relation (\ref{eq25}) reproduces exact result for 1D Ising model.

By the same token, the EOS is helpful in calculation of the entropy $S$ per spin:
\begin{equation}
S=-\frac{1}{N}\left(\frac{\partial G}{\partial T}\right)_{h,f_s}.
\label{eq26}
\end{equation}
During calculations, the contributions containing derivatives  $\left(\partial \varepsilon/\partial T\right)_{h, f_s}$ will cancel on the basis of EOS, and the final result is:
\begin{eqnarray}
S&=&k_{\rm B}\ln \left[\cosh \left(\frac{\beta h}{2}\right)+
\sqrt{\sinh^2 \left(\frac{\beta h}{2}\right) +e^{-\beta J}}\; \right] - \frac{h}{T}m\nonumber\\
&+&\frac{J}{2T}\frac{e^{-\beta J}}{\cosh \left(\frac{\beta h}{2}\right)
\sqrt{\sinh^2 \left(\frac{\beta h}{2}\right) +e^{-\beta J}}+\sinh^2 \left(\frac{\beta h}{2}\right) +e^{-\beta J}}\nonumber\\
&-&3k_{\rm B}\left \{\frac{2}{x_{\rm D}}\left[{\rm Li}_2\left(e^{-x_{\rm D}}\right)-\frac{\pi^2}{6} \right] - \ln \left(1-e^{-x_{\rm D}} \right)\right \},
\label{eq27}
\end{eqnarray}
where $J$ is the function of $\varepsilon$ obtained from the EOS.

The numerical calculations based on the above formalism will be presented in the next Section.

\section{\label{sec:results}Numerical results and discussion}

In this chapter we present the numerical results obtained for some exemplary model parameters. The choice of these parameters is such that every energy component in the total free energy is non-negligible. This enables performing a complete analysis of the model, showing the importance of every component found in EOS with a single set of model parameters.

In particular, for the exponent $n$ in the exchange integral (Eq. (\ref{eq3})) we chose the value $n=6$, which assures a quick change of $J$ vs. the distance, and is in agreement with our previous papers \cite{Balcerzak2017b}. Regarding the Morse potential, we chose the values $D/J_0=8$ for the potential depth and $\delta=5$ for its width and asymmetry. We note that $J_0$ is the strength of exchange interaction in the absence of deformation, and this parameter is useful for establishing the convenient energy scale for the studied system.
With such energy normalization, the Debye temperature in the ground state is assumed as $k_{\rm B}T_{\rm D}^0/J_0=1/3$, whereas the Gr{\"u}neisen parameter for 1D system (with the Morse potential) is taken from the exact calculations \cite{Krivtsov2011a} and amounts to $\gamma=3 \delta/2$. For the above set of model parameters, the equilibrium NN distance is found with the value $d_0/r_0=0.99821$, which results from EOS for $T=0$, $h=0$ and $f_s=0$ whereas $\varepsilon=0$. Hereafter, using the above set of constants, we present the various thermodynamic properties calculated for arbitrary temperature $T$, magnetic field $h$ and external force $f_s$. For presentation of the results we have chosen such constant external parameters (either $f_s$ or $h$) for which the dependencies of the presented quantities are most characteristic. 

\begin{figure}[h]
\begin{center}
\includegraphics[width=0.9\textwidth]{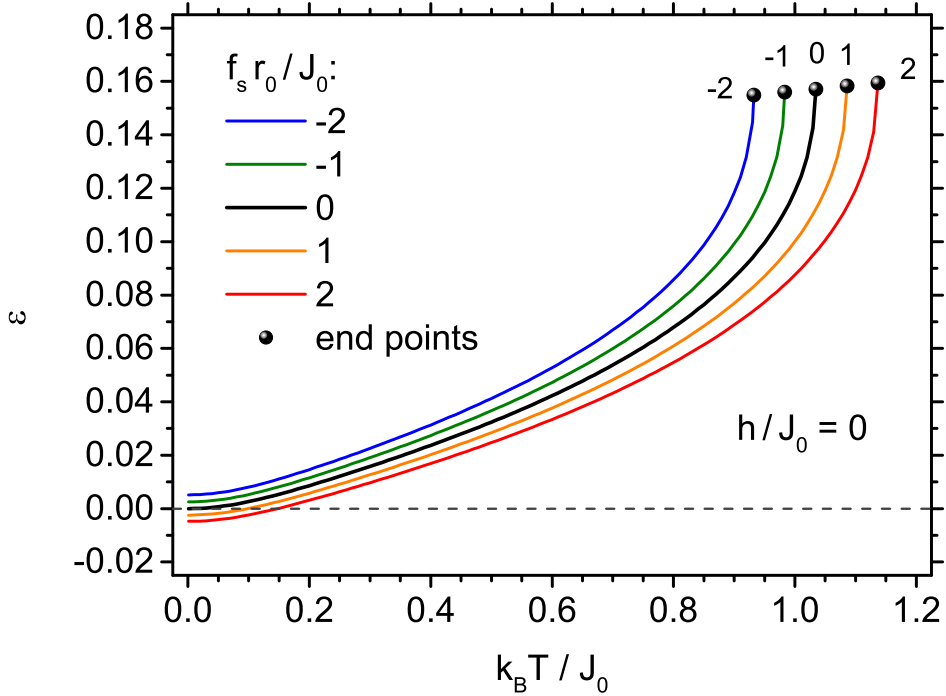}
\caption{\label{Fig1} The length deformation $\varepsilon$ vs. dimensionless temperature $k_{\rm B}T/J_0$, for $f_sr_0/J_0=-2,\,-1,\,0,\,1,\,2$ and $h/J_0=0$.} 
\end{center}
\end{figure}

\begin{figure}[h]
\begin{center}
\includegraphics[width=0.9\textwidth]{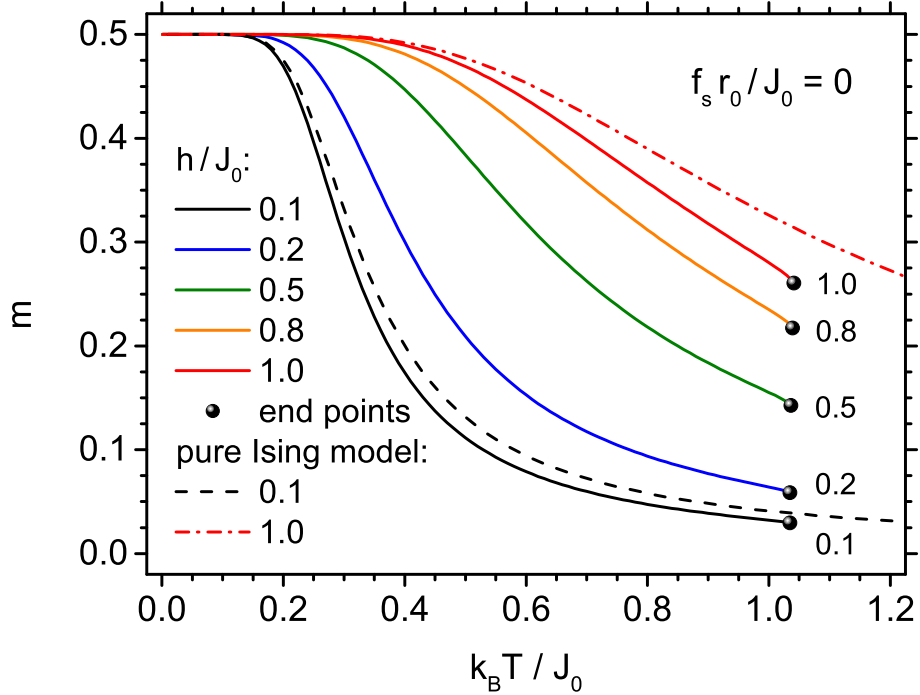}
\caption{\label{Fig2} The magnetization per spin $m$ vs. dimensionless temperature $k_{\rm B}T/J_0$, for $f_sr_0/J_0=0$ and $h/J_0=0.1,\,0.2,\,0.5,\,0.8,\,1.0$. The dashed curves present exact magnetization of the pure 1D Ising model for  $h/J_0=0.1$ and $h/J_0=1.0$, respectively.}
\end{center}
\end{figure}

\begin{figure}[h]
\begin{center}
\includegraphics[width=0.9\textwidth]{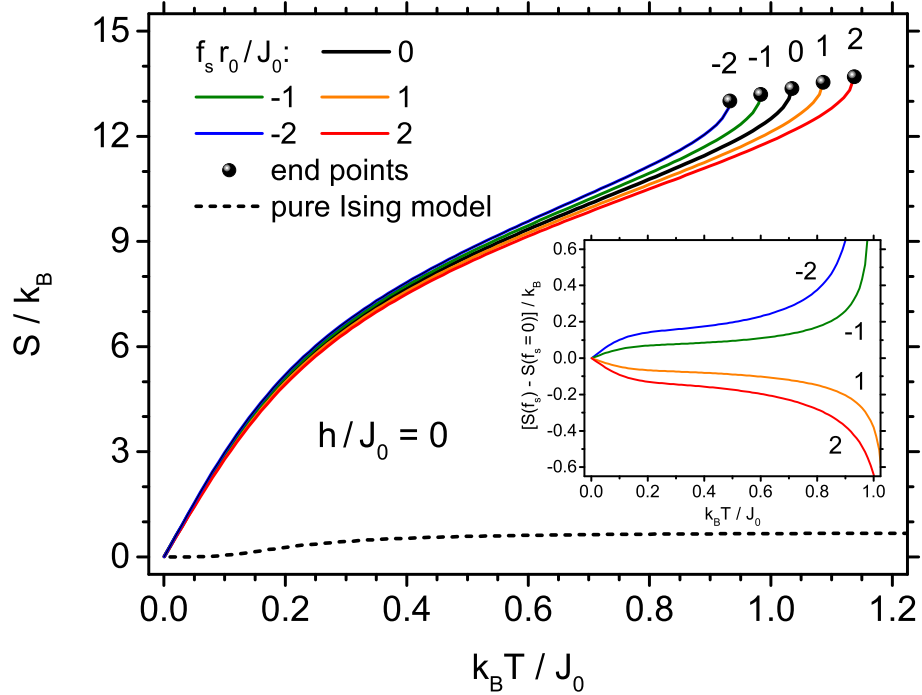}
\caption{\label{Fig3} The entropy per spin in dimensionless units, $S/k_{\rm B}$,  vs. dimensionless temperature $k_{\rm B}T/J_0$, for $f_sr_0/J_0=-2,\,-1,\,0,\,1,\,2$ and $h/J_0=0$. The dashed curve presents entropy of the pure 1D Ising model (exact result). The inset shows the difference between entropy at given $f_s$ and at $f_s=0$ as a function of the dimensionless temperature. } 
\end{center}
\end{figure}

\begin{figure}[h]
\begin{center}
\includegraphics[width=0.9\textwidth]{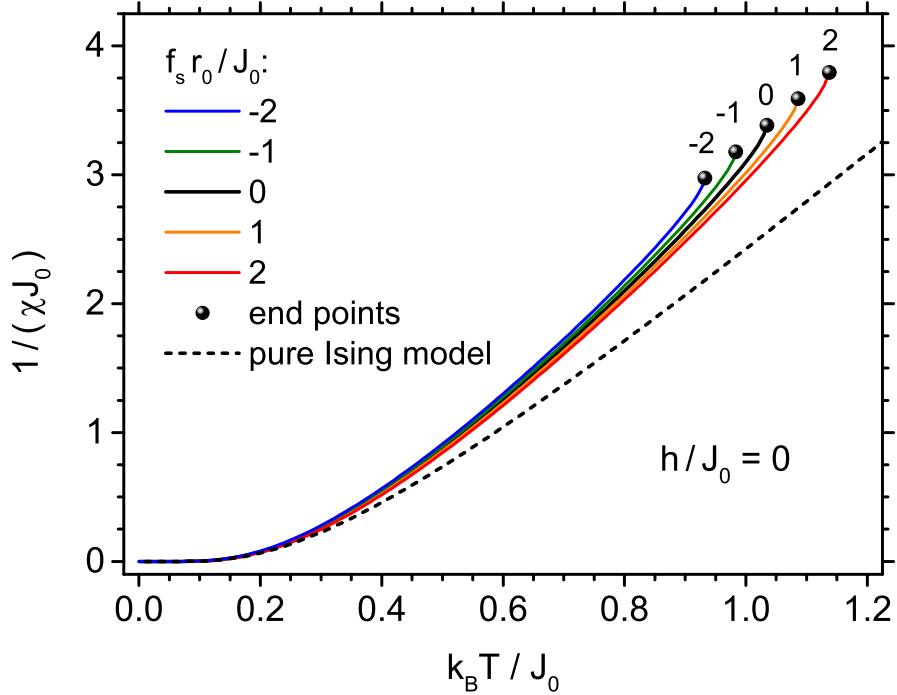}
\caption{\label{Fig4} The inverse magnetic susceptibility in dimensionless units, $1/\left(\chi J_0\right)$, vs. dimensionless temperature $k_{\rm B}T/J_0$, for $f_sr_0/J_0=-2,\,-1,\,0,\,1,\,2$ and $h/J_0=0$. The dashed curve presents inverse paramagnetic susceptibility of the pure 1D Ising model (exact result)}.
\end{center}
\end{figure}

\begin{figure}[h]
\begin{center}
\includegraphics[width=0.9\textwidth]{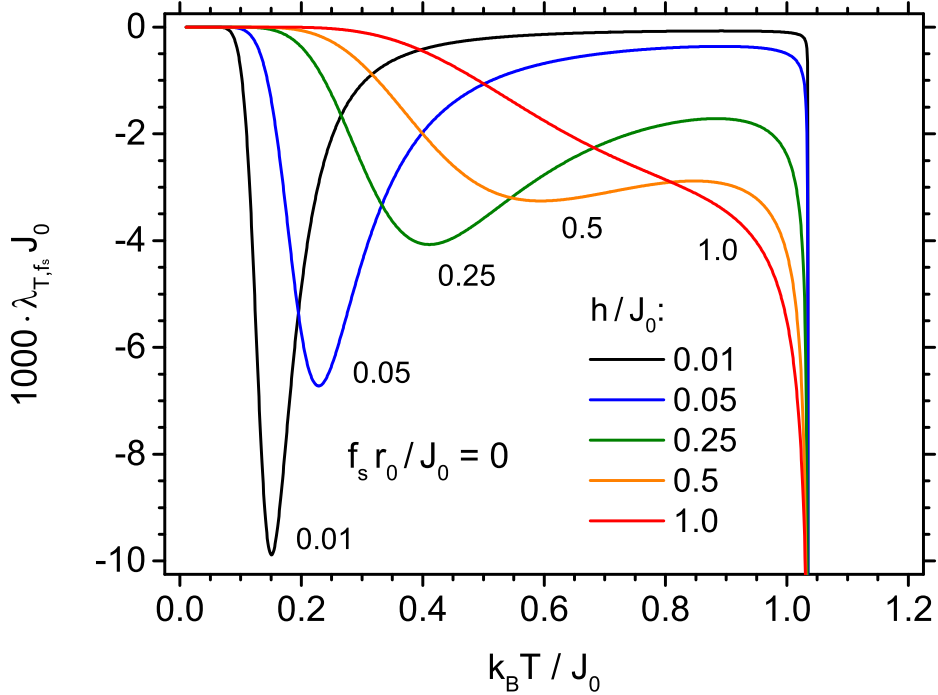}
\caption{\label{Fig5} The magnetostriction coefficient in dimensionless units, $\lambda_{T,f_s}J_0$, vs. dimensionless temperature $k_{\rm B}T/J_0$, for $f_sr_0/J_0=0$ and $h/J_0=0.01,\,0.05,\,0.25,\,0.5,\,1.0$.} 
\end{center}
\end{figure}

\begin{figure}[h]
\begin{center}
\includegraphics[width=0.9\textwidth]{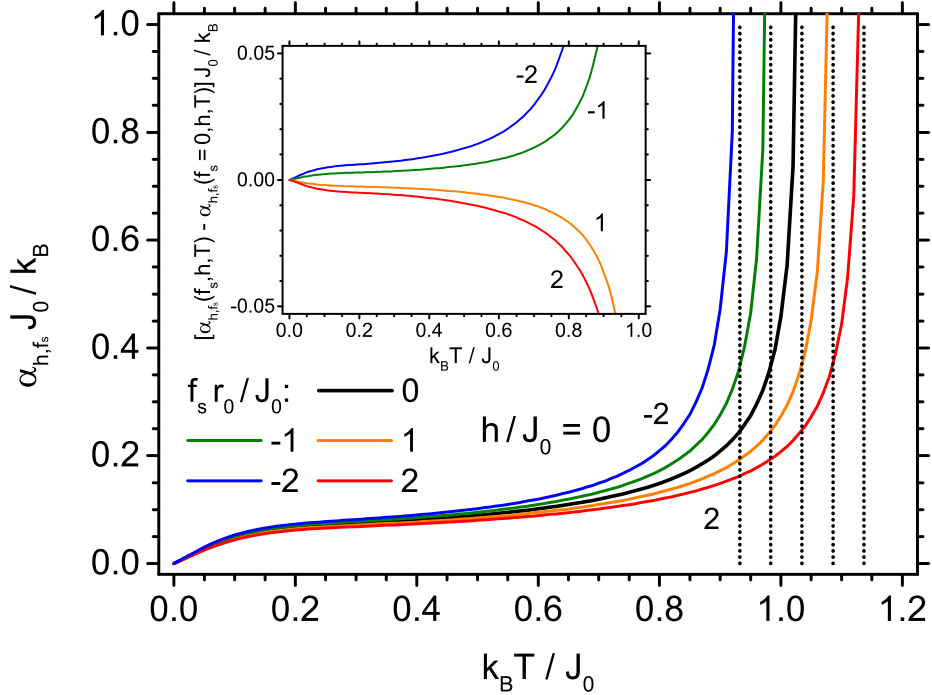}
\caption{\label{Fig6} The thermal expansion coefficient in dimensionless units, $\alpha_{h,f_s}J_0/k_{\rm B}$, vs. dimensionless temperature $k_{\rm B}T/J_0$, for $f_sr_0/J_0=-2,\,-1,\,0,\,1,\,2$ and $h/J_0=0$. The inset shows the difference between the thermal expansion coefficient at given $f_s$ and at $f_s=0$ as a function of the dimensionless temperature.}
\end{center}
\end{figure}

\begin{figure}[h]
\begin{center}
\includegraphics[width=0.9\textwidth]{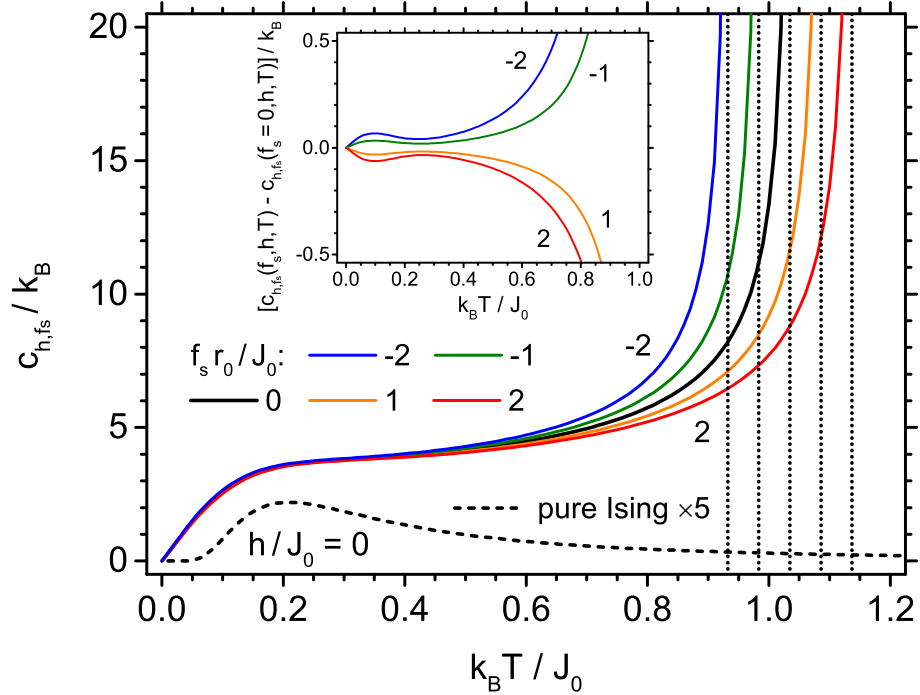}
\caption{\label{Fig7} The specific heat per lattice site in dimensionless units, $C_{h,f_s}/k_{\rm B}$, vs. dimensionless temperature $k_{\rm B}T/J_0$, for $f_sr_0/J_0=-2,\,-1,\,0,\,1,\,2$ and $h/J_0=0$.  The dashed curve presents the specific heat for the pure 1D Ising model (exact result) exhibiting a paramagnetic maximum (value of the specific heat multiplied by 5). The inset shows the difference between the specific heat at given $f_s$ and at $f_s=0$ as a function of the dimensionless temperature.} 
\end{center}
\end{figure}

\begin{figure}[h]
\begin{center}
\includegraphics[width=0.9\textwidth]{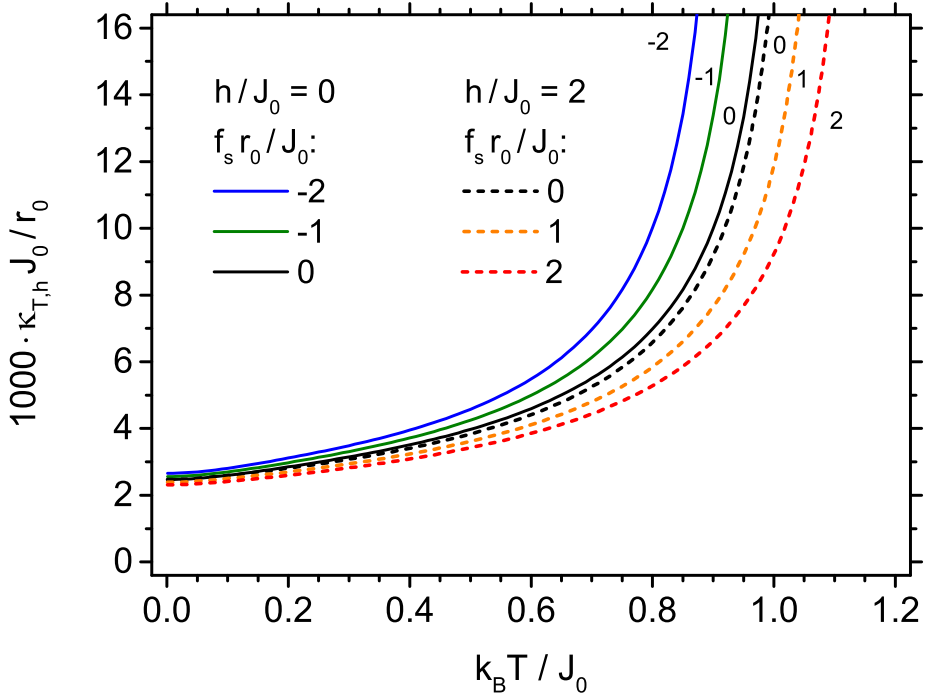}
\caption{\label{Fig8} The isothermal compressibility in dimensionless units, $\kappa_{T,h}J_0/r_0$, vs. dimensionless temperature $k_{\rm B}T/J_0$, at $h/J_0=0$ and $f_sr_0/J_0=-2,-1,0$ and at $h/J_0=2$ and $f_sr_0/J_0=0,1,2$.} 
\end{center}
\end{figure}

\begin{figure}[h]
\begin{center}
\includegraphics[width=0.9\textwidth]{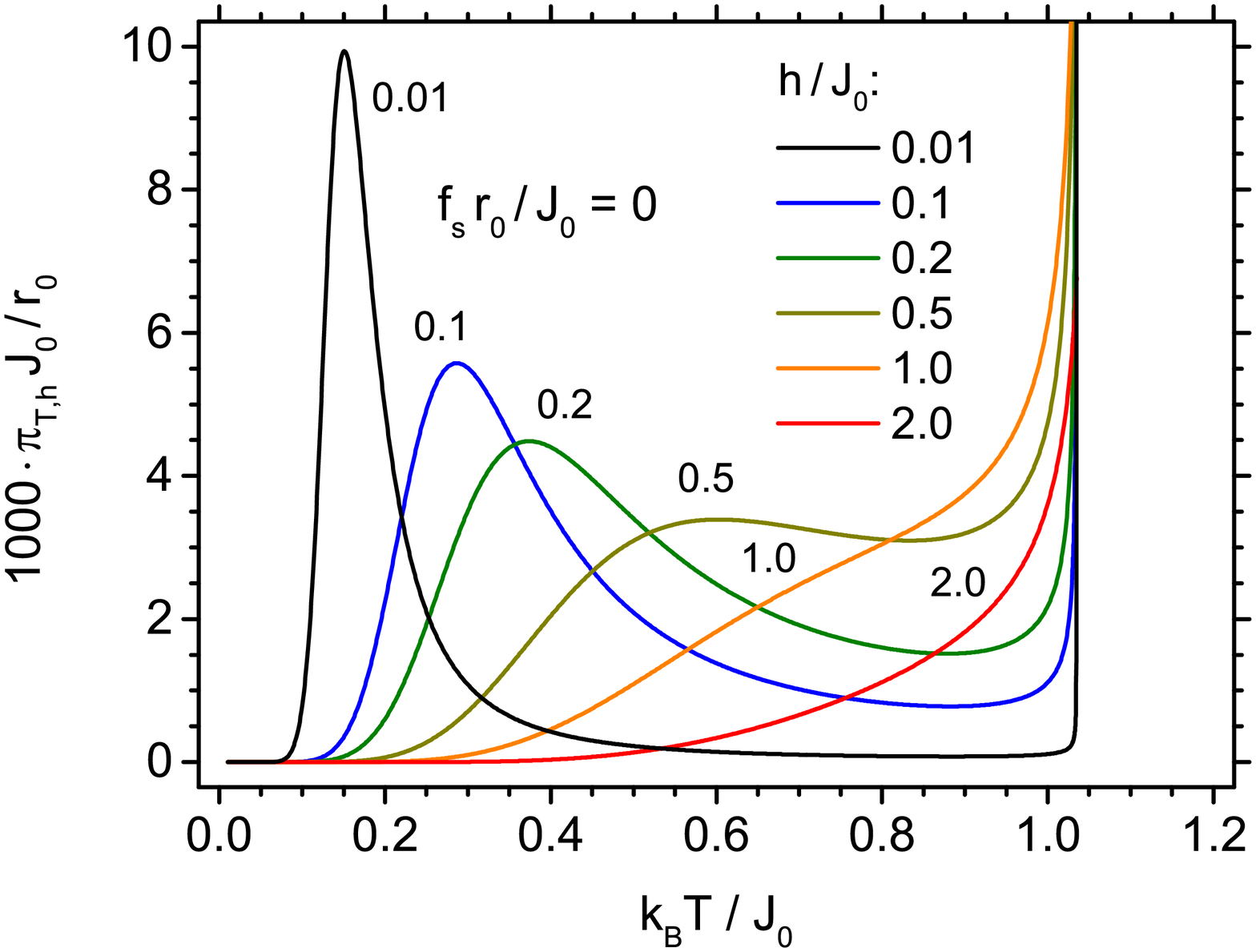}
\caption{\label{Fig9} The piezomagnetic coefficient in dimensionless units, $\pi_{T,h}J_0/r_0$, vs. dimensionless temperature $k_{\rm B}T/J_0$, for $f_sr_0/J_0=0$ and $h/J_0=0.01, 0.1,\,0.2,\,0.5,\,1.0,\,2.0$.} 
\end{center}
\end{figure}

In Fig.\ref{Fig1} the linear deformation $\varepsilon$ is presented as a function of dimensionless temperature $k_{\rm B}T/J_0$. Different curves correspond to various values of the external force $f_s$. The magnetic field is absent ($h=0$). It is seen that $\varepsilon$ is a non-linearly increasing function of $T$. For given temperature an increase in the compressive force ($f_s>0$) results in the decreasing deformation $\varepsilon$. On the other hand, when the stretching force ($f_s<0$) is increased, the deformation is enhanced. The end-points on the curves denote the limit for the range of stable solutions of EOS. Namely, for the temperature exceeding these points there is no finite solution for the length deformation and $\varepsilon \to \infty$. In our opinion this instability is connected with the melting phenomenon. It can also be seen that instability temperature strongly depends on the external force $f_s$. On the other hand, the $\varepsilon$-values at the stability end-points weakly depend on $f_s$. It can also be checked that in the absence of external forces, in the ground state (when $f_s=0$, $h=0$ and $T=0$), the deformation vanishes ($\varepsilon \to 0$), as it can be expected.

The average magnetization per spin, $m$, is presented in Fig.\ref{Fig2} as a function of temperature $k_{\rm B}T/J_0$. In this case the different curves correspond to various external magnetic fields, whereas the external force is set to zero. According to the formula Eq.(\ref{eq25}) the magnetization of 1D system vanishes for $h=0$. However, for $h>0$ the magnetization can take nonzero values and reaches the saturation value of $m=1/2$ for $T \to 0$. As one can see from Fig.\ref{Fig2}, the magnetization decreases with increasing temperature until the stability end-point for $\varepsilon$ is reached. The temperatures corresponding to the end-points only very weakly depend on $h$. At the same time, the magnetizations at the end-points can have remarkable values, depending on $h$. By the dashed curves the exact magnetization of the pure 1D Ising model is also presented for comparison. These curves are obtained when all nonmagnetic (elastic) interactions are neglected and the exchange integral remains constant.

In Fig.\ref{Fig3} the entropy per spin in $k_{\rm B}$ units, $S/k_{\rm B}$, is plotted vs. dimensionless temperature $k_{\rm B}T/J_0$ for the same parameters of $h$ and $f_s$ as in Fig.\ref{Fig1}. The entropy is defined by Eq.(\ref{eq26}) and given in the final form by Eq.(\ref{eq27}). We see that the entropy reaches zero value at $T=0$, and in the low-temperature limit its dependence on the external force vanishes. This behaviour is in agreement with the 3rd law of thermodynamics. In the region of high temperatures, for given $T$, the entropy is a decreasing function when the external force increases. This kind of  behaviour is expected for stable systems. It can also be noticed that the entropy values at the stability end-points only weakly depend on the external force $f_s$. The inset presents the difference between the entropy values at given $f_s$ and $f_s=0$, emphasizing the range of low temperatures and showing a slight asymmetry in the influence of compressive and stretching force. The entropy of the pure 1D Ising model is also plotted for comparison. This entropy is limited by the value $\ln\, 2$ when $T \to \infty$.

In Fig.\ref{Fig4} the inverse magnetic susceptibility $1/(\chi J_0)$ is drawn as a dimensionless quantity vs. temperature $k_{\rm B}T/J_0$. The susceptibility is defined as: $\chi=-\frac{1}{N}\left(\frac{\partial^2 G}{\partial h^2}\right)_{T,f_s}=\left(\frac{\partial m}{\partial h}\right)_{T,f_s}$. We note that in the numerical calculations the derivative $\left(\frac{\partial \varepsilon}{\partial h}\right)_{T,f_s}$ must be taken into account with the values resulting from the solution of EOS. In this figure the magnetic field is zero, and the external force parameters, corresponding to different curves, have the same values as in Figs.\ref{Fig1} and \ref{Fig3}. We see that the susceptibility is of paramagnetic type and its inverse only weakly depends on the external force $f_s$. However, in the high temperature region, some diminishing of 
$1/(\chi J_0)$ can be noted when the external force increases. This means an increase in the susceptibility itself, and can be explained by the fact that increasing external force makes the deformation $\varepsilon$ smaller and, as a result, the exchange integral $J\left(\varepsilon \right)$ is enhanced. It is also worth noticing that at the stability end-points the values of $1/(\chi J_0)$ are not constant but they evidently depend on the force parameter $f_s$. These changes are mainly due to the fact that, for different $f_s$, the stability end-points occur at different temperatures. For comparison, the inverse paramagnetic susceptibility of the pure 1D Ising model (exact result) is also plotted.

The magnetostriction coefficient, defined as $\lambda_{T,f_s}=\frac{1}{L}\left(\frac{\partial L}{\partial h}\right)_{T,f_s}=\frac{1}{1+\varepsilon}\left(\frac{\partial \varepsilon}{\partial h}\right)_{T,f_s}$, is plotted in dimensionless units vs. temperature in Fig.\ref{Fig5}. In this figure we assume that the external force is absent and different curves correspond to various external field values. We see that this coefficient is negative (which indicates that the magnetic force occurring in EOS is compressive) and it vanishes when $T \to 0$. For small magnetic fields a deep minimum is found in the low temperature region. This minimum results from competition of different forces in the equation of state (\ref{eq19}). We note that EOS is a strongly non-linear equation for $\varepsilon$, and the dynamics of $\varepsilon$-changes vs. $h$ and $T$ is different for various components of this equation. This results in a non-linear behaviour of $\lambda_{T,f_s}$. For higher magnetic fields the system becomes more compressed, and is more "stiff" in the low temperature region, showing a decrease in the magnetostriction magnitude. On the other hand, the system becomes very "soft" when approaching the instability temperature, where the magnetostriction diverges to $-\infty$ at the temperature weakly dependent on the magnetic field.

In Fig.\ref{Fig6} the thermal expansion coefficient, $\alpha_{h,f_s}$, is plotted  vs. temperature $k_{\rm B}T/J_0$ in dimensionless units. This coefficient is defined as $\alpha_{h,f_s}=\frac{1}{L}\left(\frac{\partial L}{\partial T}\right)_{h,f_s}=\frac{1}{1+\varepsilon}\left(\frac{\partial \varepsilon}{\partial T}\right)_{h,f_s}$. In this figure we present the results in the absence of the magnetic field ($h=0$), but for different values of the external forces $f_s$. We see that the thermal expansion coefficient diverges to $+\infty$ at the end of stability region. The vertical dashed lines indicate the positions of ending temperatures, which are in agreement with the previous figures (e.g. Figs.\ref{Fig1}, \ref{Fig3} and \ref{Fig4}). It is worth noticing that the compressive forces are decreasing the thermal expansion effect. For $T\to 0$ all curves tend to zero, which expresses a correct thermodynamic behaviour. The inset presents the difference between the thermal expansion coefficient values at given $f_s$ and $f_s=0$, which is a slightly non-monotonic function of the temperature and the influence of stretching and compressive force is slightly asymmetric.

The specific heat per atom for constant $h$ and $f_s$, $C_{h,f_s}$, is plotted vs. temperature in Fig.\ref{Fig7}. The parameters are the same as in Fig.\ref{Fig6}. The specific heat is defined by the formula: $C_{h,f_s}=T\left(\frac{\partial S}{\partial T}\right)_{h,f_s}=-\frac{1}{N}\left(\frac{\partial^2 G}{\partial T^2}\right)_{h,f_s}$, where $G$ is given by Eq.(\ref{eq17}). By comparison with Fig.\ref{Fig6} it can be noticed that the specific heat curves present qualitatively similar behaviour to the thermal expansion dependencies. In particular, for $T\to 0$ the specific heat also tends to zero, in agreement with the third law of thermodynamics. When approaching the instability point, the specific heat diverges, similarly to $\alpha_{h,f_s}$. This divergence can be expected from the analysis of the entropy vs. temperature curves near the instability point, as seen in Fig.\ref{Fig2}. The inset presents the difference between the specific heat values at given $f_s$ and $f_s=0$, emphasizing the range of low temperatures and showing a slight asymmetry in the action of the compressive ad stretching force.  
It is also worth mentioning that the paramagnetic maximum of the specific heat, which exists for the pure 1D Ising model, can be found at $k_{\rm B}T/J_0\approx 0.21$ with the value of $C_{0,0}/k_{\rm B}\approx 0.44$ (with all nonmagnetic contributions neglected). However, this smooth maximum, as seen on the dashed curve, is to weak to be noticed in the full model, when the elastic interactions and the phononic specific heat are taken into account.

The isothermal compressibility is defined by the formula $\kappa_{T,h}=-\frac{1}{L}\left(\frac{\partial L}{\partial f_s}\right)_{T,h}=-\frac{1}{1+\varepsilon}\left(\frac{\partial \varepsilon}{\partial f_s}\right)_{T,h}$, and is presented in dimensionless units vs. temperature in Fig.\ref{Fig8}. Different curves correspond to various values of both parameters $h/J_0$ and $f_sr_0/J_0$, as indicated in the figure legend. In particular, curves plotted for $h/J_0=0$ illustrate the influence of stretching force ($f_s<0$), whereas curves plotted for $h/J_0=2$ illustrate the effect of the compressive force ($f_s>0$). Moreover, comparison of the curves plotted for $f_sr_0/J_0=0$ shows the influence of the external magnetic field. It is worth noticing that the isothermal compressibility takes a non-zero value when $T\to 0$ and only very weakly depends on the external field (and force) in the ground state.

In the last figure (Fig.\ref{Fig9}) we present the piezomagnetic coefficient in dimensionless units vs. temperature. The piezomagnetic coefficient is defined by $\pi_{T,h}=\left(\frac{\partial m}{\partial f_s}\right)_{T,h}$. The curves presented in Fig.\ref{Fig9} are calculated in the absence of the external force, but for different magnetic fields $h/J_0$. We see that for $T\to 0$ the external magnetic field ($h>0$) has no influence on the magnetization, which remains there in saturated state (see also Fig.\ref{Fig2}). On the other hand, at the instability point the magnetization changes discontinuously for $h>0$, and its derivative with respect to compressive force tends to infinity. For small magnetic fields $h$, smooth maxima of $\pi_{T,h}$ are visible in the intermediate temperature region. These maxima look like the minima of magnetostriction  $\lambda_{T,f_s}$ in Fig.\ref{Fig5}. In fact, the similarity of $\pi_{T,h}$ and $-\lambda_{T,f_s}$ curves can be explained on the basis of thermodynamic Maxwell relations. If we write the differential of the total Gibbs energy as $dG=-S\,dT+L\,df_s-Nm\,dh$, then one of the Maxwell relations reads: $\left(\frac{\partial L}{\partial h}\right)_{T,f_s}=-N\left(\frac{\partial m}{\partial f_s}\right)_{T,h}$. From this formula we obtain the exact relationship between the magnetostriction and piezomagnetic coefficients in 1D system:  $\pi_{T,h}=-d_0\left(1+\varepsilon \right)\lambda_{T,f_s}$.
Since in our case $d_0 \approx r_0$ and $\varepsilon$ is small, the above equation explains why the plots shown in Figs. \ref{Fig5} and \ref{Fig9} are similar up to the sign of the plotted quantities.

\section{Summary and conclusions}

In the paper the Ising 1D chain with spin $S=1/2$ is studied within a new approach which simultaneously takes into account the elastic interactions and thermal vibrations of the atoms. Within the method the total Gibbs energy is constructed, which enables a full, self-consistent, thermodynamic description of the system. For the constant set of exemplary parameters, by means of numerical calculations we demonstrate the mutual interdependence of magnetic and mechanical properties of the system, in a wide temperature range. It can be stressed that all calculated quantities are either derivatives of the Gibbs energy or are based directly on the EOS, and they present the correct thermodynamic behaviour as a function of the temperature.
In particular, the influence of the external compressive and stretching forces, as well as the external magnetic field on the thermodynamic response functions has been demonstrated. 

In our approach two kinds of elastic energies are distinguished: vibrational and static one, both arising from the Morse potential. These energies have different magnitudes. The vibrational energy is connected with the thermal excitations. It almost vanishes at $T=0$, leaving only the quantum zero-temperature vibrations. The vibrational energy is responsible for the specific heat and in this aspect cannot be replaced by the static energy. In the extended Debye model, the vibrational energy has been taken into account in anharmonic approximation, which is especially important at high temperatures. On the other hand, the static energy is connected with deformation of the crystal with respect to some reference lattice. For instance, this energy exists when the crystal is deformed by the static external force, and is present at $T=0$ as well. The static energy can be by several orders larger than the vibrational energy. In particular, it is responsible for the isothermal compressibility, even for $T=0$, where the vibrational energy is almost negligible. Thus, in the general thermodynamic description, for simultaneous calculations of the specific heat and isothermal compressibility, both kinds of energy are irreplaceable.

It is worth noticing that in the present approach the Debye integral is calculated exactly for arbitrary temperature. Therefore, the model can be used in a full temperature range of the solid phase existence.

The coupling between magnetic and elastic properties is taken into account by the power-law dependence of the exchange integral on the distance, which is valid in the wider range of relative deformations than the linear expansion (see, for example, Ref.~\cite{Mattis}) and appears to capture better the physical situation.

The application of the Morse potential has a non-questionable advantage over the usual harmonic one. First of all, since the harmonic potential is symmetric around equilibrium positions of atoms, it is not able to describe the effect of thermal expansion. This deficiency would lead to cancellation of the results presented in Fig.1 and Fig.6, which would then be the pointless studies. Assuming the harmonic potential would also influence the rest of numerical results. In particular, the instability temperatures presented in several figures, which can be attributed to the melting phenomenon thus improving the harmonic potential-based approach, would never be reached. Moreover, the harmonic elastic potential is applicable only to the nearest-neighbour interaction, whereas the Morse potential which describes the long-range interactions, is much more physical. It should be stressed that the Morse potential has been summed up exactly for the infinite range of interaction, giving finite value of the sum per atom, therefore, no cut-off procedure is needed in 1D case. Nevertheless, it should be admitted that in the case when temperature goes to absolute zero, $T \to 0$, the harmonic potential could be used as a first approximation instead of the anharmonic one.

Therefore, the key effect of the application of the Morse potential, instead of the harmonic one, is the prediction of the instability temperature, which can be attributed to the melting phenomenon, thus improving the harmonic potential-based approach.

It is worth mentioning  that the instability temperature for solid phase, obtained here when $\varepsilon \to \infty$, should be treated as an upper limit for the melting transition. The melting temperature of real crystal is somewhat lower and should occur for finite $\varepsilon$, according to the Lindemann criterion \cite{Gilvarry_1956}. This criterion assumes that melting already occurs when the root-mean-square amplitude of thermal vibration reaches a critical fraction of the nearest-neighbour distance. However, in order to study the melting point more precisely, as a 1st-order transition between the solid and liquid phase, the Gibbs energy of liquid phase should simultaneously be at our disposal. This is not the goal of our study, aimed at describing a solid, periodically ordered magnetic system.

Regarding possibility of comparison with the experiments, let us remark that the ideal 1D system cannot be realized without its interaction with a substrate, or with a matrix medium, because the mechanical stabilization of the system is necessary. These interactions may markedly influence the properties of the linear chain, for instance, via external forces $f_s$ resulting from a difference between the thermal expansion coefficients of the chain and its substrate. Unlike in our model, such forces would be temperature dependent, and their values should be self-consistently determined by describing simultaneously the Ising chain and its coupling to the neighbourhood. However, from the theoretical point of view, it would be a much more complex model, which would prevent us, at this stage, from comparison of the ideal free chain considered in the paper with experimental situation. We think that such a problem should be addressed in the forthcoming works, especially when the experimental studies aimed at uncovering a role of elastic interactions for 1D Ising systems, for instance, studies of magnetostriction, compressibility or piezomagnetic effect, are carried out. An example of such study might be the works devoted to CoV$_2$O$_6$ \cite{Nandi2014} and CoNb$_2$O$_6$ \cite{Nandi2019} Ising chains for which magnetostriction and expansion coefficient were measured. However, it should be emphasized once more that the mentioned systems are only quasi-1D ones. 

It should be added that a characteristic feature of experimental quasi-1D Ising systems is the phase transition to the ordered state, which occurs at very low temperatures, and is caused by the weak interchain interactions. In our ideal model the phase transition is not possible at any finite temperature.

We would like to point out that the method can be extended for 1D chain with higher spins ($S\ge 1$), the long-range magnetic interactions, as well as for the quantum model, taking into account the perpendicular magnetic field. The structural disorder can also be taken into account, both for the exchange interactions, and for the elastic constants as well. As indicated in the paper \cite{Mendez-Sanchez2013}, the disorder of elastic parameters may lead to the Anderson localization in 1D system. The application of the method for 2D systems could also be of potential interest.

In our opinion, the presented method paves the way for comprehensive studies of low-dimensional magnetic solids by generalization of the thermodynamic description.

\begin{acknowledgments}

This work was partly supported within the projects VEGA 1/0531/19 and APVV-16-0186.
 
 \end{acknowledgments}


%

\newpage

\end{document}